\documentclass[12pt]{article}
\usepackage{graphicx}
\usepackage{epstopdf} 
\usepackage{amssymb} 
\usepackage{amsmath} 
\numberwithin{equation}{section}
\newtheorem{thm}{Theorem}[section]
\newtheorem{prop}[thm]{Proposition}

\DeclareMathOperator{\Dim}{Dim}
\DeclareMathOperator{\Tr}{Tr}

\begin{document}		

\centerline{\LARGE Feynman\vspace{5mm}  identity for planar graphs}

 \centerline{\large G.A.T.F.da Costa
\footnote{g.costa@ufsc.br}}
\centerline{\large Departamento
de Matem\'{a}tica} \centerline{\large Universidade Federal de
Santa Catarina} \centerline{\large
88040-900-Florian\'{o}polis-SC-Brasil}
\begin{abstract} 
The Feynman identity (FI) of a planar graph
relates
the Euler polynomial of the graph to an infinite product over the 
 equivalence classes of closed nonperiodic signed cycles in the graph. 
The main objectives of this paper are to compute the number of equivalence classes of nonperiodic cycles of given length and sign in a planar graph and to interpret the data encoded by the FI 
 in the context of free Lie superalgebras. This solves in the case of planar graphs a problem first raised by S. Sherman
and sets the FI  as the denominator identity of a free Lie superalgebra generated from a graph. Other results are obtained. For instance, in connection with zeta functions of graphs.
\vspace{5mm}

\end{abstract}

\vspace{10mm}

\section{Introduction}

 Denote by $\theta_{\pm}(N)$  the number of equivalence classes of nonperiodic cycles of length $N$ with sign $\pm 1$ in a finite  connected and non oriented planar graph $G$. Definitions are given in section 2. The univariate Feynman identity (FI) for a planar graph is the formal relation in the indeterminate $z$ that can be expressed as
\begin{equation}
 {\cal E}_{G}^{2}(z)=\prod_{N=1}^{+\infty} (1+z^{N})^{\theta_{+}(N)}
(1-z^{N})^{\theta_{-}(N)}
\end{equation}
where
\begin{equation}
 {\cal E}_{G}(z):=1+\sum_{N=1}^{|E|} a(N) z^{N},
\end{equation}
 is {\it the Euler polynomial} of $G$. The coefficient $a(N)$ is the number of subgraphs of $G$ with $N$ edges that have  all the vertices with even degree, called eulerian subgraphs.

The FI (1.1) was first conjectured by R. Feynman and proved by S. Sherman [22]. It is an 
important element of the combinatorial formalism of the Ising model in two dimensions, much studied by physicists [4].  
The FI in the general non planar case is more complicated and demands advanced ideas from algebraic topology to be understood. It was proved in full generality by M. Loebl [18] and D. Cimasoni [2]. See also the paper by T. Helmuth [9]. In the present paper, we consider the FI for planar graphs, only.

In [23] Sherman considered a special case of the multivariate version of (1.1) where the graph has a single vertex and several loops hooked to it.
He raised the problem of interpreting the FI in this special case
in algebraic terms 
after pointing out 
similarities with {\it the Witt identity}
of Lie algebra theory. The Witt identity, the $(+,-,+)$ case of relation (1.3) below,
encodes informations about the dimensions of the
vector spaces of a free Lie algebra, which are given by {\it Witt formula} (1.4), the  enveloping algebra and
 the vector space that generates the algebra. See [21]:

\begin{prop} \label{pro1} 
If ${\cal V}$ is an $R$-dimensional vector space and $L$ is the free Lie algebra generated by  ${\cal V}$, then 
$L= \bigoplus_{N=1}^{\infty} L_{N}$, $L_{N}$ has dimension given by
${\mathcal M}(N;R)$, and
\begin{equation} \label{(1.2)}
\prod_{N=1}^{\infty} (1-z^{N})^{\pm {\mathcal M}(N;R)} = 1\mp \sum_{N=1}^{\infty} {f}_{\pm}(N) z^{N},
\end{equation}
\begin{equation}
{\mathcal M}(N;R)=\frac{1}{N} \sum_{g | N} \mu(g) R^{\frac{N}{g}},
\end{equation}
where $f_{+}(1)=R$, $f_{+}(N)=0$, if $N>1$, $f_{-}(N)=R^{N}$. The summation is over all divisors of $N$ and $\mu$ is the M\"obius function: $\mu(+1)=1$, $\mu(g)=0$, if $g=p_{1}^{a_{1}}...p_{q}^{a_{q}}$ with $a_{i}>1$, and $\mu(p_{1}...p_{q})=(-1)^{q}$, $p_{1}$, ..., $p_{q}$ primes. In the $(-,+,-)$ case, (1.3) is the generating
function for the dimensions of the homogeneous subspaces of the enveloping algebra of $L$. 
\end{prop}

In a series of papers S.-J.
Kang and collaborators
generalized Proposition 1.1 to more general free Lie algebras and superalgebras [11-13].  In this more general setting
Sherman's problem for the special case of (1.1) was solved in
[5,6] .   
Now, we turn our attention to (1.1) for general planar graphs. 
This is now possible thanks to the formulation of
the FI using the {\it edge adjacency matrix} [24] and the 
{\it Kac-Ward transition matrix} [2] of a planar graph. Using these matrices and  ideas from [5] and [16] we derive counting formulas for the numbers $\theta_{\pm}(N)$ in (1.1). Then, 
using these counting formulas as {\it generalized Witt formulas} and results from [7] and [11-13] we solve Sherman's problem.

The paper is organized as follows. In section 2, the edge adjacency matrix and the Kac-Ward transition matrix of a planar graph are defined and used to obtain 
counting formulas for the numbers of signed cycles (Theorem 2.1) and of equivalence classes of nonperiodic cycles (the $\theta_{\pm}$ in (1.1)) (Theorem 2.2).
 In section 3, the results are encapsulated in three theorems and two remarks. In Theorem 3.1, which is important for section 4, other forms of the FI are given. The other theorems and remarks are complimentary or of general nature.
Section 4 has two subsections. In the first one,  we define a zeta function associated with the Kac-Ward transition matrix and show how it relates to the Ihara zeta function of a graph.  
In the second subsection,  the results in previous sections and subsection are put together with results from [11-13]. These are collected in two propositions. They are the base of our interpretation of the FI (1.1) as the denominator (or generalized Witt) identity of a Lie superalgebra 
 generated by a general finite connected  planar graph and the signed cycles in them.
In a previous paper [7], inspired by Sherman's problem and the results in [5,6], we have considered  the same problem in connection with identities which involve non-signed cycles in general graphs. They are associated with the Ihara and Bowen-Lanford zeta functions. New interpretations of these functions aroused from relating them with Lie algebras. Some results in that paper will be needed in section 4.

Let's add three final comments to this introduction. The combinatorial formalism of the Ising model was invented to avoid the complications of the algebraic formalism based on Lie algebras used by L. Onsager [20]  in the calculation of the exact partition function for the two dimensional Ising model.
The present paper shows that there is a connection, which deserves further investigation, of  the combinatorial formalism of the Ising model with Lie algebraic ideas.
The results on the Sherman's problem in [5,6] and in the present paper
 establishes a connection going from graph theoretical ideas to  some  of the foundational results in [11-14]. However,
the results  can be understood
as a graph theoretical representation 
of some  results in [11-14]. In the case of non planar graphs, the right hand side of (1.1) becomes a finite sum of infinite products. It is possible to rewrite it in terms of a single infinite product and then use the same ideas of [11-13] to interpret the result algebraically, as in section 4.

\section{Basic definitions and cycle counting formulas}

In this section, we define cycles and the graph matrices needed to derive the counting formulas given in Theorems 2.1 and 2.2.

Call $G=(V,E)$ a finite  connected and non oriented planar graph, $V$ is the set of
vertices with $|V|$ elements and $E$ is the set of non oriented edges with $|E|$ elements  labelled $e_{1}$,
...,$e_{|E|}$. The
graph may have multiple edges and loops but no 1-degree vertices. 
Consider the graph $G'$ built from $G$ by fixing an orientation for the edges of $G$ and adding
in the opposing oriented edges $e_{|E|+1}=(e_{1})^{-1}$,
...,$e_{2 |E|}=(e_{|E|})^{-1}$, $(e_{i})^{-1}$ being the oriented edge
opposite to $e_{i}$ and with origin (end) the end (origin) of
$e_{i}$. In the case that $e_{i}$ is an oriented loop,
$e_{i+|E|}=(e_{i})^{-1}$ is just an additional oriented loop hooked
to the same vertex. Thus, $G'$  has $2|E|$ oriented
edges. An edge with vertices $v_{i}$ and $v_{j}$ with the orientation from $v_{i}$ to $v_{j}$ is said to have origin at $v_{i}$ and end at $v_{j}$.
A path in $G$ is given by an ordered sequence of edges $(e_{i_{1}},...,e_{i_{N}})$, $i_{k} \in \{1, ..., 2|E|\}$,  in  $G'$ such that the end of $e_{i_{k}}$ is the origin of $e_{i_{k+1}}$.

In this paper we call a cycle a closed path, that is, the end of $e_{i_{N}}$ coincides with the origin of $e_{i_{1}}$,
 subjected to the non-backtracking condition that $e_{i_{k+1}}  \neq e_{i_{k}+|E|}$. In another words,
a  cycle  never goes immediately backwards over a  previous edge. 
The length of a cycle is the number of edges in its sequence. A cycle $p$ is called periodic if $p=q^r$ for some $r>1$ and subsequence $q$. If there is no such $r>1$, the cycle is called non periodic. Number $r$ is called the period of $p$. The cycle $(e_{i_{N}}, e_{i_{1}}, ...,e_{i_{N-1}})$ is called a circular permutation of $(e_{i_{1}},...,e_{i_{N}})$
and $(e_{i_{N}}^{-1},...,e_{i_{1}}^{-1})$ is an inversion of the latter. 
The circular permutations of a sequence represent the same cycle $p$, hence, they constitute an equivalence class denoted by $[p]$. Equivalent cycles have the same length. We will consider a cycle and its inversion as distinct. This is the reason for the square on the left hand side of (1.1). 
The sign of a cycle $p$ is given by the formula 
\begin{equation} \label{a}
s(p)=(-1)^{1+n(p)}
\end{equation}
where 
$n(p)$ is the number of integral revolutions of a vector tangent  to  $p$. Equivalent cycles have the same sign. A cycle and the inverse cycle have the same sign. (The sign of a cycle $p$ can be obtained drawing a normal curve  compatible with the cycle. Then, Whitney's theorem says that the sign is given by $(-1)^{V_{0}}$ where $V_{0}$ is the number of self-intersections of the curve. See [3]).

In order to count cycles of a given length and sign in a non oriented graph $G$ we need the {\it edge adjacency matrix} [24] and {\it the  Kac-Ward transition matrix} of $G$ [2].
The edge adjacency matrix is the $2|E| \times 2|E|$ matrix $T(G)$ with entries indexed by the edges of $G'$ 
defined by
\begin{equation} \label{b}
  {T(G)}_{e,e'} =  \left \{ \begin{array}{ll}
                             1  & \mbox{ if $f(e)=s(e')$ but $e' \neq e^{-1}$;}\\
                              $0$ & \mbox{otherwise,}
                                  \end{array}
                    \right.
\end{equation}
where $f(e)$ is the end vertex of edge $e$ and $s(e')$ is the  vertex at the origin of $e'$.
The transition matrix of $G$  
 is the $2|E|\times 2|E|$ matrix $S(G)$ with entries also indexed by the edges of $G'$ and
given by
\begin{equation} \label{c}
  {S(G)}_{e,e'} =  \left \{ \begin{array}{ll}
                             exp( \frac{i}{2} \alpha(e,e'))  & \mbox{ if $f(e)=s(e')$ but $e' \neq e^{-1}$;}\\
                              $0$ & \mbox{otherwise.}
                                  \end{array}
                    \right.
\end{equation}
Fix an orienting frame on ${\mathbb R}^{2}$ and a unit circle with center at the origin $(0,0)$. Define the turning angle $\alpha(e,e')$ as the net number of rotations
around the unit circle made by a vector equivalent to the unit tangent vector to the edge $e$  as it starts, for instance, from the middle of this edge and ends at the middle point of $e'$. The angle is positive for anticlockwise rotations and negative for clockwise ones.
The structure of matrix $S$, except for the complexity of the entries, resembles very much that of $T$ given in [10] and [24] and can be obtained similarly. The matrix can be expressed as 
\begin{equation*}
 \left( \begin{array}{clcr}
A & B\\
C & D 
\end{array} \right)
\end{equation*}
where  $A$, $B$, $C$ and $D$ are $|E| \times |E|$ matrices.
Denote by $\overline{M^{t}}$ the transpose conjugate of $M$.
Given the graph $G'$ with the edges labeled as in the introduction,
a) $S_{e,e'}=0=S_{e',e}$ if $e'=e^{-1}$ , by definition of $S$, b) $B=\overline{B^{t}}$, c)
$C=\overline{C^{t}}$. The diagonal entries of $B$ and $C$ are zero. d) $D=\overline{A^{t}}$. The diagonals of $A$ and $D$ are zero if the graph has no loops.

Theorems 2.1 and 2.2 below give counting formulas for the number of cycles with given sign and length  in a planar graph in terms of powers of the matrices $T$ and $S$. The counting formulas for the exponents $\theta_{\pm}$ in (1.1) are given in Theorem 2.2.

\begin{thm} \label{b2} Given a graph $G$,
denote by ${\cal K}_{\pm}(N)$ the number of cycles  with sign $\pm 1$. Then, 
\begin{equation} \label{c1}
\Tr T^{N} = {\cal K}_{+}(N)+ {\cal K}_{-}(N),
\end{equation}
\begin{equation} \label{d}
 \Tr S^{N} = {\cal K}_{-}(N)- {\cal K}_{+}(N),
\end{equation}
\begin{equation} \label{e}
{\cal K}_{\pm}(N)=\frac{1}{2} \Tr \left[ T^{N} \mp S^{N} \right].
\end{equation}
\end{thm}

\noindent{\bf Proof.} 
 Let $a$ and $b$ be two edges of $G$. The $(a,b)^{th}$ entry of
matrix $T^{N}$ is
\begin{equation*}
(T^{N})_{(a,b)}= \sum_{{e}_{{i}_{1}}, ..., {e}_{{i}_{N-1}}} {T}_{(a,
{e}_{{i}_{1}})} {T}_{({e}_{{i}_{1}},{e}_{{i}_{2}})}...{T}_{({e}_{{i}_{N-1}},b)}.
\end{equation*}
The definition  of $T$ gives that
$(T^{N})_{(a,b)}$ counts the number of paths of length $N$ with no backtracks from
edge $a$ to  edge $b$. For $b=a$, only closed paths are
counted. Taking the trace gives the number of cycles with
every edge taken into account as starting edge, hence, the trace overcounts cycles because every edge in the cycle is taken into account as starting edge.
The cycles counted by the trace are tail-less, that is, $e_{i_{1}}  \neq e_{i_{N}}^{-1}$; otherwise,  $\Tr T^{N}= \sum_{a} (T^{N})_{(a,a)}$ would have a term with entry $(a,a^{-1})$ which is not possible. From the definition of  ${\cal K}_{\pm}(N)$, relation (2.4) follows.  On the other hand,
the $(a,b)^{th}$ entry of
matrix $S^{N}$ is
\begin{equation*}
(S^{N})_{(a,b)}= \sum_{{e}_{{i}_{1}}, ..., {e}_{{i}_{N-1}}} {S}_{(a,{e}_{{i}_{1}})} {S}_{({e}_{{i}_{2}},{e}_{{i}_{3}})}...{S}_{({e}_{{i}_{N-1}},b)}.
\end{equation*}
From (2.3) it follows that
for $b=a$  each non zero product in the summand equals 
\begin{equation*}
e^{i n \pi}=(-1)^{n}=(-1) sign(p),
\end{equation*}
for some integer $n$ and cycle $p$ of length $N$. 
 Hence, taking the trace gives the total number of positive signs which is equal to the number of  positive cycles, ${\cal K}_{+}(N)$, times $-1$, plus the total number of negative signs which is equal to $-{\cal K}_{-}(N)$, times $-1$,
and we get (2.5).
From this and (2.4), (2.6) will  follow.
 $\Box$

In [16] M. Lin investigates combinatorial aspects of a multivariate special case of (1.1) when the graph has one vertex and several loops hooked to it. We apply  ideas from [16] to prove relations (2.7) and (2.8) below.

\begin{thm} \label{b3}
Denote by $\theta(N)$  the number of equivalence classes of  non periodic cycles of length $N$ in a graph $G$ and by
$\theta_{\pm}(N)$  the number of equivalence classes of  non periodic cycles of length $N$ with sign $\pm 1$, $\theta(N)=\theta_{+}(N)+\theta_{-}(N)$. Then,
\begin{eqnarray}
\theta_{+}(N) &=& \frac{1}{N}\sum_{g \hspace{1mm} odd \mid N}\mu(g) {\cal
K}_{+} \left( \frac{N}{g} \right),\\
 \theta_{-}(N) &=& \frac{1}{N}\sum_{g
\hspace{1mm} even  \mid N}\mu(g) {\cal K}_{+} \left( \frac{N}{g}
\right)+ \frac{1}{N}\sum_{ g  \mid N} \mu(g){\cal K}_{-}
\left( \frac{N}{g} \right),\\
\theta(N) & = &  \frac{1}{N}\sum_{g \hspace{1mm} \mid N}\mu(g) 
Tr T^{ \frac{N}{g}}.
\end{eqnarray}
\end{thm}
where ${\cal K}_{\pm}$ is given by (2.6).

\noindent{\bf Proof.} See [16], section 2.2. The sign of a cycle of period $g$, $p=(h)^{g}$,  can be expressed as 
\begin{equation*}
sign(p) = (-1)^{g+1} (sign(h))^{g}.
\end{equation*}
For $g$ odd, $sign(p)=+1$ if and only if $sign(h)=+1$, hence,
\begin{equation}
{\cal K}_{+}
(N)
=\sum_{g \hspace{1mm} odd \mid N} \frac{N}{g} \theta_{+} \left(
\frac{N}{g} \right).
\end{equation}
Inverting this relation gives (2.7). The proof is similar to the one of M\"obius inversion formula  [15]. 
Now, $sign(p)=-1$ whenever $g$ is even or, for any $g$, we have that $sign(h)=-1$.  Therefore, 

\begin{equation}
{\cal K}_{-} (N) = \sum_{g \hspace{1mm}
even \mid N} \frac{N}{g} \theta_{+} \left(\frac{N}{g} \right) +
\sum_{ g|N } \frac{N}{g} \theta_{-} \left(
\frac{N}{g} \right),
\end{equation}
and
\begin{equation}
{\cal K}(N):= {\cal K}_{+}(N) + {\cal K}_{-}(N)=\sum_{ g
\mid N} \frac{N}{g} \theta \left( \frac{N}{g} \right).
\end{equation}
By M\"obius inversion,
\begin{equation*}
\theta(N)=\theta_{+}(N)+\theta_{-}(N)= \frac{1}{N}\sum_{ g \mid N}
\mu(g){\cal K} \left( \frac{N}{g} \right).
\end{equation*}
Using (2.6), we get (2.9). Besides,
\begin{eqnarray*}
\theta_{-}(N) &=& \frac{1}{N} \sum_{g \mid N} \mu(g) {\cal K}\left( \frac{N}{g} \right) - \theta_{+}(N)\\
& = & \frac{1}{N} \sum_{ g \mid N} \mu(g) \left(
{\cal
K}_{+}\left( \frac{N}{g} \right) + {\cal K}_{-}\left( \frac{N}{g} \right) \right) - \frac{1}{N} \sum_{g \hspace{1mm}  odd \mid N}\mu(g)
{\cal K}_{+}\left( \frac{N}{g} \right)\\
& = & \frac{1}{N} \sum_{g \hspace{1mm} even \mid N}\mu(g) {\cal
K}_{+}\left( \frac{N}{g} \right) + \frac{1}{N}\sum_{ g \mid N} \mu(g){\cal K}_{-}\left( \frac{N}{g} \right).
\end{eqnarray*}
$\Box$

\section{Other forms of the FI}

In this section we derive some identities which
connect the FI (1.1) to an exponential, a determinant, formal Taylor expansions and an explicit formula for the coefficients. These other forms of the FI are given in Theorem 3.1 below. Together with Theorem 2.2, Theorem 3.1  is  important for the algebraic interpretation of  the FI  in section 4. The determinantal identity (3.3) below is called the {\it Kac-Ward formula} and it is well known in the literature about the Ising model. 
See [2], [17], and references therein, for other derivations of this formula. 
The meaning of $\Omega(N)$ in (3.3) will be clear from Theorem 3.2.  
In Theorem 3.3 we collect several recursions relating the coefficients $a(N)$ in (1.2), the Taylor expansions (3.2) and the exponents $\Omega(N)$. Remarks 3.1-2 give complimentary results.

\begin{thm}  \label{b5}
Define 
\begin{equation}
g(z):= \sum_{N=1}^{+\infty} \frac{\Tr S^{N} }{N} z^{N}, 
\end{equation}
Then,
\begin{eqnarray}
{\cal E}_{G}^{\pm 2}(z)
& = & e^{\mp g(z)}= 1 \mp  \sum_{i=1}^{+\infty} {c}_{\pm}(i) z^{i}\\
& = & \prod_{N=1}^{+\infty} (1-z^{N})^{\pm \Omega (N)}
= [\det \left( I-z S \right)]^{\pm 1},
\end{eqnarray}
where
\begin{equation}
\Omega(N):= \frac{1}{N} \sum_{g | N} \mu(g) \Tr S^{ \frac{N}{g}}
\end{equation}
and
\begin{equation} \label{3.5}
{c}_{\pm}(i)= \sum_{m=1}^{i} \lambda_{\pm}(m) \sum_{
\begin{array}{l} {a}_{1}+2{a}_{2}+...+i{a}_{i} =i\\
{a}_{1}+...+{a}_{i} = m \end{array}} 
 \prod_{k=1}^{i} 
\frac{(   \Tr S^{k}    )^{{a}_{k}}}{{a}_{k}! k^{{a}_{k}}}
\end{equation}
where $\lambda_{+}(m)=(-1)^{m+1}$, $\lambda_{-}(m)=+1$.
 Furthermore,
\begin{equation} \label{3.6}
   \Tr S^{N} = N  \sum_{
\begin{array}{l}  s = ({s}_{i})_{i \geq 1}, {s}_{i} \in {\bf Z}_{\geq 0}\\
 \sum i{s}_{i}=N  \end{array}} (\pm 1)^{|s|+1}
 \frac{(\mid s \mid -1)!}{s!} \prod {c}_{\pm}(i)^{{s}_{i}}.
\end{equation}
where $|s|=\sum s_{i}$, $s!=\prod s_{i}!$.
\end{thm}
\noindent{\bf Proof.} See [5], [16] and [11-13]. Take the formal logarithm of both sides of (1.1) to get
\begin{eqnarray*}
ln {\cal E}_{G}^{\pm 2}(z)
& = & \pm \sum_{N'=1}^{+\infty} \left[ \theta_{+}(N') ln (1+z^{N'}) + \theta_{-}(N') ln(1-z^{N'}) \right] \\
& = & \pm \sum_{N'=1}^{+\infty}  \left[ \theta_{+}(N') \sum_{l=1}^{+\infty} (-1)^{l-1}
\frac{z^{N'l}}{l} + \theta_{-}(N') (-1) \sum_{l=1}^{+\infty} \frac{z^{lN'}}{l}\right]\\
&=& \pm \sum_{N'=1}^{+\infty}  \sum_{l=1}^{+\infty} \left[ (-1)^{l-1} \theta_{+}(N')-\theta_{-}(N') \right]
\frac{z^{N'l}}{l}\\
&=& \mp \sum_{N'=1}^{+\infty}  \sum_{l=1}^{+\infty} \left[ (-1)^{l} \theta_{+}(N')+\theta_{-}(N') \right]
\frac{z^{N'l}}{l}\\
&=& \mp \sum_{N=1}^{+\infty} {\cal L}(N)z^{N},
\end{eqnarray*}
where
\begin{equation*}
{\cal L}(N) :=
\sum_{g|N} \frac{1}{g}  \left[     (-1)^{g} \theta_{+}
\left( \frac{N}{g} \right)
+\theta_{-} \left( \frac{N}{g} \right) \right].
\end{equation*}
Decompose  ${\cal L}(N)$ as a sum over the even divisors of $N$ plus a sum over the odd divisors of $N$.
Using $\theta=\theta_{+}+\theta_{-}$ it results that
\begin{eqnarray*}
 -{\cal L}(N) & =  & - \sum_{g \hspace{1mm} even \mid N} \frac{1}{g} \theta \left( \frac{N}{g} \right)
+
  \sum_{g \hspace{1mm} odd \mid N} \frac{1}{g} \left[\theta_{+} \left( \frac{N}{g} \right)
- \theta_{-} \left( \frac{N}{g} \right) \right]\\
&=& -\frac{1}{N} \sum_{ g|N} \frac{N}{g} \theta \left( \frac{N}{g} \right)
+\frac{1}{N} \sum_{g \hspace{1mm} odd |N} \frac{N}{g} 2 \theta_{+} \left( \frac{N}{g} \right).
\end{eqnarray*}
Using (2.12) and (2.6),
\begin{equation*}
-N{\cal L}(N)= -{\cal K}(N)+2{\cal K}_{+}(N)={\cal K}_{+}(N)-{\cal K}_{-}(N)=\Tr(-S^{N}),
\end{equation*}
so that
\begin{equation*}
{\cal L}(N)= \frac{Tr S^{N}}{N}.
\end{equation*}
By Jacobi trace formula,
\begin{eqnarray*}
ln {\cal E}_{G}^{\pm 2}(z)
&=& \pm \sum_{N=1}^{+\infty} \frac{\Tr (-S^{N})}{N} z^{N} 
= \pm \Tr ln(1-zS)
= \pm ln \det(1-zS),\\
{\cal E}_{G}^{\pm 2}(z) &=& [det \left( I-z S \right)]^{\pm}.
\end{eqnarray*}
Set
\begin{equation*}
\Omega(N)= \sum_{g | N} \frac{\mu(g)}{g} {\cal L} \left( \frac{N}{g} \right)
\end{equation*}
so
\begin{equation*}
{\cal L}(N)= \sum_{g | N} \frac{1}{g} \Omega \left( \frac{N}{g} \right).
\end{equation*}
Then,
\begin{align*}
ln {\cal E}_{G}^{\pm 2}(z)&= \mp \sum_{N=1}^{+\infty} \sum_{g | N} \frac{1}{g} \Omega \left( \frac{N}{g} \right) z^{N} = \pm \sum_{N} \Omega(N) ln(1-z^{N}),\\
\end{align*}
and we get (3.3).

The coefficients  $c_{\pm}(i)$ are given by
\begin{equation*}
{c}_{\pm}(i) =
\frac{1}{i!}
\frac{d^{i}}{d z^{i}} \left[(1-e^{ \mp g}) \right]|_{z=0}.
\end{equation*}
Using Faa di Bruno's formula as in [5],  the derivatives  can be computed explicitly to give (3.5).

To prove (3.6) write
\begin{equation*}
\sum_{k=1}^{+\infty} \frac{\Tr S^{k}}{k} z^{k} =\mp ln \left( 1 \mp \sum_{i=1}^{+\infty} {c}_{\pm}(i) z^{i} \right) 
= \mp \sum_{l=1}^{+\infty} \frac{(-1)}{l} \left( \pm \sum_{i} {c}_{\pm}(i) z^{i} \right)^{l}.
\end{equation*}
Expand the right hand side in powers of $z$ to get:
\begin{align*}
  \sum_{k=1}^{+\infty}    z^{k} 
 \sum_{\begin{array}{l}  s = ({s}_{i})_{i \geq 1}, {s}_{i} \in {\bf Z}_{\geq 0}\\
 \sum i{s}_{i}=k \end{array}} 
 (\pm 1)^{|s|+1} \frac{(\mid s \mid -1)!}{s!} \prod c_{\pm}(i)^{s_{i}}.
\end{align*}
Comparing the coefficients give the result. $\Box$

The meaning of $\Omega(N)$ becomes clear by the following result:

\begin{thm} \label{b4}
\begin{equation}
 \Omega(N)= \left\{ \begin{array}{ll}
\theta_{-}(N)-\theta_{+}(N) & \mbox{if $N$ is odd} \\
\theta_{-}(N)-\theta_{+}(N)+\theta_{+}(\frac{N}{2}) & \mbox{if $N$ is even}
\end{array}
\right.
\end{equation}
\end{thm}

\noindent{\bf Proof.} From (2.7) and (2.8),

\begin{eqnarray*}
\theta_{+}(N) -\theta_{-}(N) &=& 
\frac{1}{N}\sum_{g \hspace{1mm} odd \mid N}\mu(g) {\cal
K}_{+} \left(\frac{N}{g} \right)
 - \frac{1}{N}\sum_{g
\hspace{1mm} even  \mid N}\mu(g) {\cal K}_{+}\left(\frac{N}{g}\right) \\
& -&  \frac{1}{N}\sum_{ g  \mid N} \mu(g){\cal K}_{-}
\left( \frac{N}{g} \right)\\
& =& 
\frac{1}{N}\sum_{g \hspace{1mm} odd \mid N}\mu(g) \left[{\cal
K}_{+} \left( \frac{N}{g} \right)- {\cal K}_{-}
\left( \frac{N}{g} \right) \right] \\
& -&  \frac{1}{N}\sum_{g
\hspace{1mm} even  \mid N}\mu(g) \left[{\cal K}_{+} \left( \frac{N}{g}
\right)+ {\cal K}_{-}
\left( \frac{N}{g} \right)\right].\\
\end{eqnarray*}
The sum over the odd divisors of $N$ equals 
\begin{equation*}
  \frac{1}{N}\sum_{ g
\mid N}\mu(g) \Tr (-S^{\frac{N}{g}})
- \frac{1}{N}\sum_{g \hspace{1mm} even \mid N}\mu(g)
\left[{\cal K}_{+} \left( \frac{N}{g}
\right)- {\cal K}_{-}
\left( \frac{N}{g} \right)\right].
\end{equation*}
We get
\begin{eqnarray*}
\theta_{+}(N) -\theta_{-}(N) &=& 
 -\Omega(N)
-\frac{2}{N}\sum_{g \hspace{1mm} even \mid N}\mu(g)
{\cal K}_{+} \left( \frac{N}{g}
\right).\\
\end{eqnarray*}
Thus, $\Omega(N)=\theta_{+}(N) -\theta_{-}(N)$, if $N$ is odd. If $N$ is even,
the even divisors of  $N=2^{j} n$ are $2^{k}$, $k=1,...,j$, and $2^{i}p$, $i=1,2,...,j$, and $p$ are the odd divisors of $n$. However, $\mu=0$ for the cases $k,i \geq 2$ , hence, using that $\mu(2p) =\mu(2) \mu(p)=-\mu(p)$, we get that
\begin{eqnarray*}
\frac{2}{N}\sum_{g \hspace{1mm} even \mid N}\mu(g)
{\cal K}_{+} \left( \frac{N}{g}
\right)&=&
 \frac{2}{N}\sum_{p
\hspace{1mm} odd  \mid   N }\mu(2p)        {K}_{+} \left( \frac{N}{2p} \right)\\
&=&
-\frac{2}{N}\sum_{p
\hspace{1mm} odd  \mid   N/2 }\mu(p)        {K}_{+} \left( \frac{N}{2p} \right)\\
&=& -\theta_{+} \left(\frac{N}{2} \right)
\end{eqnarray*}
 $\Box$

\noindent {\bf Remark 3.1.} There are graphs with the property that $\Omega(N)=0$ for all $N \geq N_{0}$, for some $N_{0}$. In another words, $\theta_{+}(N)=\theta_{-}(N)$, for all odd  $N \geq N_{0}$,  and $\theta_{-}(N)=\theta_{+}(N)-\theta_{+}(N/2)$, for all even $N \geq N_{0}$. This is the case of the graph with one vertex and $R$ loops hooked to it. This is proved in reference [5]. A simpler proof is as follows.
From Theorem 3.1, relation (3.2) and (1.1), 
\begin{equation*}
-2z \frac{d}{dz} ln{\mathcal E}=\sum_{N \geq 1} Tr S^{N} z^{N}.
\end{equation*}
 Using that ${\mathcal E}(z)=(1+z)^R$,
\begin{equation*}
-2z \frac{d}{dz} ln{\mathcal E}=\sum_{N \geq 1} (-1)^{N}2R z^{N},
\end{equation*}
we get $\Tr S^{N}=(-1)^{N}2R$. Then, by (3.4), $\Omega(1)=-2R$, $\Omega(2)=2R$, $\Omega(N)=0$, if $N \geq 3$, and
\begin{equation*}
\prod_{N=1}^{+\infty} (1-z^{N})^{ \Omega (N)}= (1-z)^{-2R} (1-z^2)^{2R}= (1+z)^{2R}.
\end{equation*}
Another example is provided by the graph obtained from $R$ copies of the graph with vertices and two edges linking them, glued at the vertices. It has one vertex of degree 2 at the far left and another one at the far right and $R-2$ vertices of degree 4 in between them.
The Euler polynomial  is ${\mathcal E}(z)=(1+z^{2})^{R}$ so
\begin{equation*}
-2z \frac{d}{dz} ln{\mathcal E}=\sum_{N \geq 1} 
(-1)^{N}4Rz^{N}.
\end{equation*}
We get $\Tr S^{N}=(-1)^{N/2}4R$, if $N$ is even, and $\Tr S^{N}=0$, if $N$ is odd, so that $\Omega(N)=0$, if $N$ is odd,  $\Omega(2)=-2R$,   $\Omega(4)=+2R$, $\Omega(N)=0$, if $N$ is even and $N \geq 6$, so
\begin{equation*}
\prod_{N=1}^{+\infty} (1-z^{N})^{ \Omega (N)}= (1-z^{2})^{-2R} (1-z^4)^{2R}= (1+z^{2})^{2R}.
\end{equation*}

\noindent{\bf Remark 3.2.} 
By the Schur decomposition method there is a matrix $P$ and an upper triangular matrix $J$ with the eigenvalues $\lambda_{i}$ of ${S}$ along the diagonal such that $ S = P J P^{-1}$, hence,
\begin{equation*}
\Tr {{ S}}^{N} = \Tr ( P J P^{-1} )^{N} = \Tr J^{N} = \sum_{i=1}^{2|E|} \lambda_{i}^{N},
\end{equation*}
and
\begin{equation*}
 {\Omega}(N)= \frac{1}{N} \sum_{g |N} \mu(g) \Tr { S}^{\frac{N}{g}} =  \sum_{i=1}^{2|E|}
\frac{1}{N} \sum_{g |N} \mu(g) \lambda_{i}^{\frac{N}{g}} 
= \sum_{i=1}^{2|E|} {\mathcal M}(N;\lambda_{i}),
\end{equation*}
where ${\mathcal M}$ is given in (1.4). Using Witt identity (see (1.3), case $(+,-,+)$),
\begin{equation*} 
\det(1-z{ S})=\prod_{N=1}^{+\infty} (1-z^{N})^{ {\Omega}(N)}= \prod_{i=1}^{2|E|} (1-\lambda_{i}z).
\end{equation*}
The zeros of ${\mathcal E}_{G}^{2}(z)$, as a complex function in $z$, are the reciprocals of the eingevalues of $S$.
Expanding the product,
\begin{equation*} 
\det(1-z S)=\prod_{i=1}^{2|E|} (1-\lambda_{i}z)= 1-\Tr S z+ \dots + \det S 
z^{2|E|}.
\end{equation*}
This polynomial is the square of the Euler polynomial $1+\sum_{i=1}^{|E|} a(i)z^{i}$ so that $-\Tr{S}=2a(1)$ and $\det S=a^{2}(|E|)$, where $a(i)$ is the number of eulerian subgraphs with $i$ edges, hence, $-\Tr S \geq 0$ and $-\Tr S$ is the number of loops in the graph so that $\sum_{i} \lambda_{i}=0$ if the graph $G$ has no loops and $\sum_{i} \lambda_{i}<0$, otherwise. Also, $\det S \geq 0$ and the square root of
 $\det S$ is the number of eulerian graphs with $|E|$ edges, hence, $\det S=0$ if and only if the graph is not itself eulerian and $\det S=1$, otherwise. We may conclude that 
$S$ has  zero as an eigenvalue if and only if the graph itself is not eulerian.

\begin{thm} \label{b6} Set $\omega(n):= \Tr { S}^{n}$. Then,
\begin{align} 
{c}_{\pm} (1) &= \omega(1)=\Omega(1), \label{j1}\\
 n {c}_{\pm} (n) &=  \omega(n) \mp \sum_{k=1}^{n-1}  \omega(n-k) {c}_{\pm}(k), n \geq 2,\label{j2}\\
 {c}_{-}(n) &= {c}_{+}(n)+\sum_{i=1}^{n-1} {c}_{+}(i) {c}_{-}(n-i), n \geq 2, \label{j3}\\
 |{c}_{+}(n)| & \leq   {c}_{-}(n), \label{j4}\\
 \Omega(n) &= {c}_{+}(n)+\frac{1}{n} \sum_{k=1}^{n-1}\left( \sum_{g \mid k}
 g \Omega (g)\right) {c}_{+}(n-k) 
 - \sum_{n \neq g \mid n}
 \frac{g}{n} \Omega(g) \label{j5},\\
 2n a(n)&=-n {c}_{+}(n)+ \sum_{k=1}^{n} (3k-n)a(k){c}_{+}(n-k).
\end{align}
\end{thm}

\noindent {\bf Proof.} See reference [7].

\section{The FI and Lie superalgebras}

\subsection{The zeta functions $\zeta_{I}$ and $\zeta_{KW}$}

In this subsection  additional results are  remarked. Some of them will be relevant in the next subsection.
For instance, we define a zeta function $\zeta_{KW}$ associated with the Kac-Ward transition matrix $S$ and show how it relates to the Ihara zeta function of a graph and derive two product identities. We give examples of graphs with the same $\zeta_{KW}$.

\noindent {\bf Remark 4.1.} 
 Calculations similar to those in Theorem 3.1 show that
\begin{equation}
\prod_{N=1}^{+\infty} (1-z^{N})^{\theta(N)}=\det(I-zT).
\end{equation}
See [7,24]. The reciprocal of this relation is known in association with the {\it Ihara zeta function} $\zeta_{I}(z)$ of a graph: $\zeta_{I}(z)=\det(1-zT)^{-1}$. Therefore,  it seems  natural to define  $\zeta_{KW}(z)=\det(1-zS)^{-1}$.
 In subsection 4.2, this function will be associated to an algebra and the dimensions of its subspaces. Define
\begin{equation}
{g}_{\pm}(z)=\sum_{N=1}^{+\infty} \frac{{\cal K}_{\pm}(N)}{N} z^{N}.
\end{equation}
Then,
\begin{equation}
\zeta_{I}(z)=e^{2{g}_{+}(z)} \zeta_{KW}(z), \hspace{5mm} \zeta_{I}(z) \zeta_{KW}(z)=e^{2{g}_{-}(z)} .
\end{equation}
From these two relations one can get the identities (3.3) and (4.1).
Also,
\begin{equation}
 \prod_{N=1}^{+\infty} \left( \frac{1+z^{N}}{1-z^{N}} \right)^{\theta_{+}(N)}
=\frac{\det(1-z S)}{\det(1-z T)},
\end{equation}
\begin{equation}
 \prod_{N=1}^{+\infty} \left( \frac{1+z^{N}}{1-z^{N}} \right)^{\theta_{-}(N)}
=\frac{\det(1-z^{2} T)}{\det(1-z T) \det(1-zS)}.
\end{equation}
Other identities are possible to obtain.

\noindent{\bf Remark 4.2.}
Two non isomorphic graphs can have the same Ihara zeta function $\zeta_{I}(z)$. Also,
two non isomorphic graphs can have the same Euler polynomial (see [1,8,19], and references therein), hence, the same $\zeta_{KW}(z)$. For instance,
the graphs in Figure 1 of [8]. One of them is the graph in Remark (...), second example, with$R=3$. The other one is the graph with 6 edges which has a circle subgraph with 4 vertices but a pair of the consecutive vertices has extra two edges linking them.
They have the same Euler polynomial ${\mathcal E}(z) = (1+z^2)^3$.
From Remark 3.1, $\Tr S^{N}=0$, if $N$ is odd, and $\Tr S^{N}=12 (-1)^{N/2}$, if $N$ is even, so $\Omega(N)=0$, if $N$ is odd,  $\Omega(2)=-6$,   $\Omega(4)=+6$, $\Omega(N)=0$, if $N$ is even and $N \geq 6$. Both graphs have the same sequence
$\{\Omega (N), N \geq 1\}$.
It follows from the recursion relation (3.12)
 that two graphs with same $\zeta_{KW}(z)$ will have in common the same sequence $\{\Omega (N), N \geq 1\}$, and the same form (3.3) of the FI.

\subsection{ Feynman meets Lie}

In this subsection the FI (1.1) is associated to a free Lie superalgebra, using  results obtained by
 S.-J. Kang and collaborators in [11-14]. They generalized Proposition 1.1 (see section 1) to more general free Lie (super)algebras generated by infinite graded vector spaces. 
 We apply  their results to give a new interpretation of the FI (1.1) and of the associated function $\zeta_{KW}(z)$. The results in this subsection   can be understood 
as examples arising from graph theoretical ideas
of some  results in [11-14].

The results from [11-13] which are relevant for our objectives are summarized
in the Propositions 4.1 and 4.2 below. Relations (4.6-7) and (4.11-13) are called {\it the generalized Witt formulas}; relations (4.8) and (4.14), the $(+,-,+)$ cases, are called {\it the denominator  or generalized Witt identities} of the algebras.

\begin{prop} \label{prA}
 Let ${\cal V}= \bigoplus_{N=1}^{\infty}
{\cal V}_{N}$ be a ${\mathbb{Z}}_{>0}$-graded superspace  with finite dimensions $\dim {\cal V}_{N}= |t(N)|$ and superdimensions
$\Dim {\cal V}_{N}= t(N) \in {\mathbb{Z}}$, $\forall i \geq 1$. 
Let ${\mathcal L}=
\bigoplus_{N=1}^{\infty} {\mathcal L}_{N}$ be the free Lie superalgebra generated
by ${\cal V}$ with a ${\mathbb{Z}}_{>0}$-gradation induced by that of ${\cal V}$. Then, the ${\mathcal L}_{N}$
superdimension  is
\begin{equation} \label{lie1}
\Dim {\mathcal L}_{N}= \sum_{g | N} \frac{\mu(g)}{g}  W \left(\frac{N}{g}\right).
\end{equation}
The summation ranges over all positive divisors $g$ of $N$ and $W$ is given by
\begin{equation} \label{lie2}
 W(N)= \sum_{s \in T(N)} \frac{(\mid s \mid -1)!}{s!} \prod t(i)^{s_{i}},
\end{equation}
where $T(N)=\{ s = (s_{i})_{i \geq 1} \mid s_{i} \in {\mathbb{Z}}_{\geq 0}, 
\sum is_{i}=N \}$
and $\mid s \mid = \sum s_{i}$, $ s! = \prod s_{i} !$. Furthermore,
\begin{equation}\label{lie3}
\prod_{N=1}^{\infty} (1-z^{N})^{\pm \Dim {\mathcal L}_{N}}= 1 \mp \sum_{N=1}^{\infty} {f}_{\pm}(N) z^{N},
\end{equation}
with $f_{+}(N)=t(N)$ and $f_{-}(N)=\Dim U({\mathcal L})_{N}$,
where $\Dim U({\mathcal L})_{N}$ is the dimension of the $N$-th homogeneous subspace of the
 universal enveloping algebra $U({\mathcal L})$ and the generating function for the $W$'s,
\begin{equation}\label{lie4}
g(z) :=\sum_{N=1}^{\infty} W(N)z^{N},
\end{equation}
satisfies
\begin{equation}\label{lie5}
e^{-g(z)}=  1-\sum_{N=1}^{\infty} t(N) z^{N}.
\end{equation}
\end{prop}
$\Box$

See section 2.3 of [12]. Given a formal power series $\sum_{N=1}^{+\infty} t_{N} z^{N}$ with $ t_{N} \in {\mathbb Z}$, for all $i \geq 1$,
the coefficients in the series can be interpreted as the superdimensions of a ${\mathbb Z}_{>0}$-graded  superspace ${\cal V}= \bigoplus_{i=1}^{\infty}
{\cal V}_{N}$ with dimensions $dim {\cal V}_{N}= |t_{N}|$ and superdimensions
$\Dim {\cal V}_{N}= t_{N} \in {\mathbb Z}$.
Let ${\mathcal L}$ be the free Lie superalgebra generated by ${\cal V}$. Then, it has a gradation induced by ${\cal V}$ and its homogeneous subspaces have dimensions given by (4.6) and (4.7). 
Let's consider the (+) case of  (3.3).
Apply the previous interpretation to $\det(1-z  S)$ as a polynomial of degree $2|E|$ in the formal variable $z$. This is a power series with coefficients $t_{N}=-c_{+}(N)$, if $N \leq 2|E|$, and $t_{N}=0$, if $N > 2|E|$.
Comparison of the relations
in Theorem 3.1 with those
in Proposition 4.1 yields:
given a graph $G$, $S$ its associated Kac-Ward transition matrix, 
let
${\cal V}= \bigoplus_{N=1}^{2|E|} 
{\cal V}_{N}$ be a ${\mathbb{Z}}_{>0}$-graded superspace with finite dimensions 
$\dim {\cal V}_{N}= |c_{+}(N)|$ and the superdimensions
$\Dim {\cal V}_{N}= -c_{+}(N)$ where $-c_{+}(N)$ is the coefficient of $z^{N}$ in $\det(1-zS)$. Let ${\mathcal L}=
\bigoplus_{N=1}^{\infty} {\mathcal L}_{N}$ be the free Lie superalgebra generated
by ${\cal V}$. Then, the  ${\mathcal L}_{N}$
superdimension is  $\Dim{\mathcal L}_{N} =\Omega(N)$ and $\zeta_{KW}(z)$ is the generating function for the dimensions of the subspaces of the enveloping algebra of ${\mathcal L}$. These can be computed recursively using the recursions in Theorem 3.3.
If we raise both sides of the plus case of  (3.3)  to $1/2$ its right hand side is the generating function of eulerian subgraphs, so in this case the vector space ${\cal V}_{N}$ is generated by the eulerian subgraphs of size $N$.
In [7] we  have already applied Proposition 4.1 to give an algebraic  interpretation of (2.9) and (4.1).

Another interpretation follows from the next proposition:

\begin{prop} {\it
Let ${\cal V}= \bigoplus_{(n,a) \in {\mathbb Z}_{>0} \times {\mathbb Z}_{2}} {\cal V}_{(n,a)}$ be a
${\bf ({\mathbb Z}_{>0} \times {\mathbb Z}_{2}) }$-graded  colored superspace
with superdimensions $\Dim  {\cal V}_{(n,a)}=t(n,a) \in {\mathbb Z}$, $\forall (n,a) \in {\mathbb Z}_{>0} \times {\mathbb Z}_{2}$.
Let
${\cal L}= \bigoplus_{(n,a) \in {\mathbb Z}_{>0} \times {\mathbb Z}_{2} }
{\cal L}_{(n,a) }$ be the free Lie superalgebra
generated by $V$.
Then, the dimensions of the homogeneous subspaces ${\cal L}_{(n,a) }$
are given by 
\begin{equation}
\Dim {\cal L}_{(n,0)}=
\sum_{g | n } \frac{\mu(g)}{g} W \left(\frac{n}{g},0\right)
+
\sum_{g \hspace{1mm} even | n } \frac{\mu(g)}{g} W \left(\frac{n}{g},1\right),
\end{equation}
and
\begin{equation}
\Dim {\cal L}_{(n,1)}
= \sum_{g \hspace{1mm} odd | n } \frac{\mu(g)}{g} W \left(\frac{n}{g},1\right),
\end{equation}
where
\begin{equation}
W(\tau,b) = \sum_{s \in T(\tau,b)} \frac{(|s|-1)}{s!}
\prod t(\tau_{i},b_{j})^{s_{ij}}
\end{equation}
and 
\begin{equation*}
T(\tau,b)=\{ s = ({s}_{i,j})_{i,j \geq 1} \mid {s}_{i,j} \in {\mathbb{Z}}_{\geq 0}, 
\sum {s}_{i,j}(\tau_{i},{b}_{j})=(\tau,b) \},
\end{equation*}
which is the set of partitions of $(\tau,b)$ into a sum of $(\tau_{i},b_{j})$'s, $\mid s \mid = \sum s_{i,j}$, $ s! = \prod s_{i,j} !$.
Furthermore,
\begin{equation}
\prod_{(n,a) \in {\mathbb Z}_{>0} \times {\mathbb Z}_{2}} (1-E^{(n,a)})^{\pm \Dim {\cal L}_{(n,a)}}
= 1 \mp {T}_{{\mathbb Z}_{>0} \times {\mathbb Z}_{2}}^{\pm},
\end{equation}
where the $E^{(n,a)}$ are basis elements of $\mathbb{C}[{\mathbb Z}_{>0} \times {\mathbb Z}_{2}]$, $E^{(n,a)} E^{(m,b)}=E^{(n+m,a+b)}$,
\begin{equation}
{T}_{{\mathbb Z}_{>0} \times {\mathbb Z}_{2}}^{\pm} := \sum_{(n,a)\in {\mathbb Z}_{>0} \times {\mathbb Z}_{2}} {f}_{\pm}(n,a)
E^{(n,a)}
\end{equation}
$f_{+}(n,a)=t(n,a)$, and $f_{-}(n,a)=\Dim  {\mathcal U}({\mathcal L})_{(n,a)}$ is the superdimension of the homogeneous subspace $(n,a)$ of the enveloping algebra
${\mathcal U(L)}$.
The generating function for the $W$'s,
\begin{equation}
g(z) :=\sum_{ (\tau,a) \in  {\mathbb Z}_{>0} \times {\mathbb Z}_{2} } W(\tau,a)E^{(n,a)},
\end{equation}
satisfies
\begin{equation}
e^{-g}=  1-{T}_{{\mathbb Z}_{>0} \times {\mathbb Z}_{2}}.
\end{equation}
}
\end{prop}
$\Box$

On the base of Proposition 4.2 we interpret the data defined on a graph in terms of the data in this proposition. First, let's
make the specialization $E^{(n,0)}=z^{n}$ and $E^{(n,1)}=z^{n}q$ with $q^{2}=1$. It follows that
\begin{equation*}
\prod_{n=1}^{+\infty} (1-z^{n})^{\Dim {\cal L}_{(n,0)}}(1-qz^{n})^{\Dim {\cal L}_{(n,1)}}
= 1- \sum_{n=1}^{+\infty} (t(n,0)+qt(n,1))z^{n}.
\end{equation*}
In particular, for $q=-1$, we get
\begin{equation*}
\prod_{n=1}^{+\infty} (1-z^{n})^{\Dim {\cal L}_{(n,0)}}(1+z^{n})^{\Dim {\cal L}_{(n,1)}}
= 1- \sum_{n=1}^{+\infty} (t(n,0)-t(n,1))z^{n},
\end{equation*}
which has the same form of the FI,
and, for $q=1$, we get
\begin{equation*}
\prod_{n=1}^{+\infty} (1-z^{n})^{\Dim {\cal L}_{(n,0)}+\Dim {\cal L}_{(n,1)}}
= 1- \sum_{n=1}^{+\infty} (t(n,0)+t(n,1))z^{n}.
\end{equation*}
Set 
\begin{equation}
t'(n):=t(n,0)-t(n,1), \hspace{5mm} t(n):=t(n,0)+t(n,1).
\end{equation}
In the case $q=1$ define ${\cal V}_{n}=\bigoplus_{a} {\cal V}_{(n,a)}$
and ${\cal L}_{n}=\bigoplus_{a} {\cal L}_{(n,a)}$. Then, ${\cal V}= \bigoplus_{n=1}^{\infty}
{\cal V}_{n}$ 
with dimensions given by
$t(n)= \sum_{a} t(n,a)=t(n,0)+t(n,1)$
becomes a graded vector space and the free superalgebra ${\cal L}$ on ${\cal V}$ has a gradation ${\cal L}= \bigoplus_{n }
{\cal L}_{n}$ induced by ${\cal V}$ with dimensions given by (4.6) and (4.7).
Therefore, one gets the data in the Proposition 4.1.

In order to fix our algebraic interpretation of the counting formulas (2.7) and (2.8) we need to know the dimensions of the spaces ${\cal V}(n,a)$. This information  comes from the data from a graph given by the matrices $T$ and $S$ as follows.
Suppose one knows $t'(n)$ and $t(n)$ but not $t(n,0)$ and $t(n,1)$.
In this case, $t(n,0)$ and $t(n,1)$ can be computed using
\begin{equation}
t(n,0)=\frac{1}{2}(t'(n)+t(n)), \hspace{5mm} t(n,1)=\frac{1}{2}(t(n)-t'(n))
\end{equation}
which follow from (4.18). The numbers $t(n)$ and $t'(n)$ are given by the coefficients of the non constant terms in $\det(1-zT)$ and $\det(1-zS)$, respectively. The numbers $t(n) \pm t'(n)$ are always even integers as proved next.

\begin{thm}
The coefficients in the polynomials in $z$ given by
\begin{equation}
\det(1-zT) \pm \det(1-zS)
\end{equation}
are even integers.
\end{thm}

\noindent{\bf Proof.} Using (4.4),
\begin{equation*}
\det(1-zT) \pm \det(1-zS)=\left[1\pm \prod_{N \geq 1}(1+2\sum_{k=1}^{+\infty} z^{Nk})^{\theta_{+}(N)  }  \right]
\det(1-zT).
\end{equation*}
Furthermore, putting $z^{N}=z'$,
\begin{equation*}
(1+2\sum_{k=1}^{+\infty} z'^{k} )^{\theta_{+}(N)} = \sum_{m \geq 0}
{\alpha}_{m} z'^{m},
\end{equation*}
where $\alpha_{0}=1$ and
\begin{equation*}
{\alpha}_{m}= \frac{2}{m} \sum_{k=1}^{m}(k\theta_{+}(N)-m+k){\alpha}_{m-k}.
\end{equation*}
 $\Box$

With the specialization $q=-1$,
the data in Proposition 4.2  together with relations (2.7) and (2.8) in Theorem 2.2 and those in Theorem 3.1 yields:
given a graph $G$ with edge and transition matrices $T$ and $S$, respectively, let
 ${\cal V}= \bigoplus_{(n,i) \in {\mathbb Z}_{>0} \times Z_{2}} {\cal V}_{(n,i)}$ be a
${\bf (Z_{>0} \times Z_{2}) }$-graded  colored superspace with
superdimension
$$
Dim  {\cal V}_{(n,i)}=t(n,i):= \frac{1}{2}(t'(a)+(-1)^{i}t(a))
$$
where the $t'$ are given by the coefficients of $\det(1-zS)$ and $t$ by  the coefficients of $\det (1-zT)$.
Let
${\cal L}= \bigoplus_{(n,i) \in Z_{>0} \times Z_{2} }
{\cal L}_{(\alpha,i) }$ be the free Lie superalgebra
generated by ${\cal V}$.
Then, the dimensions of the homogeneous subspaces ${\cal L}_{(\alpha,i) }$
are given by (2.7) and (2.8), that is,
$Dim {\cal L}_{(n,0)}= \theta_{-}(n)$
and
$Dim {\cal L}_{(n,1)}
= \theta_{+}(n)$,
and these satisfy the FI which plays the role of the $(+,-,+)$ case of (4.14).
The generating function for the dimensions of the subspaces of
the enveloping algebra of ${\mathcal L}$ is given by $\zeta_{KW}(z)$, the $(-,+,-)$ case of (4.14).

 \noindent{\bf Remark 4.3.} Define the supermatrix
\begin{equation}
 Q=\left( \begin{array}{clcr}
S & 0\\
0 & T 
\end{array} \right)
\end{equation}
and the supertrace $Str Q=\Tr S - \Tr T$ so that 
\begin{equation}
\theta_{+}(N) = \frac{1}{2N}\sum_{g \hspace{1mm} odd \mid N}\mu(g)  Str Q^{ \frac{N}{g} }
\end{equation}
Then, the quotient of the two determinants in (4.4) can be expressed as the superdeterminant  (the Berezinian)  $Ber(1-zQ)$.
Using the superformalism one can make a connection with the algebras in [14] which sugests a concrete link of these algebras with graph theoretical ideas.

\subsection*{Acknowledgments}
 Many thanks to Prof. A. Goodall (Charles University, Prague) and Prof. K. Markstr\"om (Umea University, Sweden)  for sending me references [8] and [1,19], respectively. Special thanks to Prof.
Asteroide Santana (UFSC) for help with latex commands.

\noindent {\bf References}

\noindent 1.   Andr\'en, D.,  Markstr\"om, K.: A bivariate Ising polynomial of a graph. Discrete Appl. Math. {\bf 157}, 2515-2524 (2009).

\noindent  2.   Cimasoni, C.: A generalized Kac-Ward formula,
 J. Stat. Mech.  page P07023, 2010.
 
\noindent 3. da Costa, G. A. T. F.: Feynman identity:a special case, J. Math. Phys. {\bf 38}, 1014-1034 (1997). 

\noindent 4. da Costa, G. A. T. F.,  Maciel, A. L.: Combinatorial formulation of Ising model revisited.
Rev. Brasil. do Ensino de F\'{i}sica {\bf 25}, 49-61 (2003).

\noindent 5.  da Costa, G. A. T. F.,  Variane, J.:
 Feynman identity: a special case
revisited.
Letters in Math. Phys.  {\bf 73},  221-235  (2005).
 
\noindent 6. da Costa, G. A. T. F., Zimmermann, G. A.:
 An analog to the Witt identity.
Pacific Journal of Mathematics, {\bf 263}, 475-494 (2013).

\noindent 7. da Costa, G. A. T. F.: Graphs and generalized Witt identities.  arXiv:1409.5767v2 [math.CO] (2014).

\noindent 8. Garijo, D., Goodall, A., Nesetril,J.:  Distinguishing graphs by their left and right homomorphism profiles. European J. Combin. {\bf 32}, 1025-1053 (2011).

\noindent 9.  Helmuth, T.: Ising model observables and non-backtracking walks.
J. Math. Phys. 55, 1-28 (2014).

\noindent 10.   Horton, M. D.: Ihara zeta functions of irregular graphs. Doctoral dissertation, University of California, CA (USA) 2006.

\noindent 11.  Kang, S.-J.,  Kim,  M.-H.: Free Lie algebras, generalized Witt formula, and the denominator identity.
J. Algebra {\bf 183},  560-594 (1996).

\noindent 12. Kang, S.-J.:
Graded Lie superalgebras and the superdimension formula. J. Algebra {\bf 204},  597-655 (1998).

\noindent 13.   Kang, S.-J.,  Kim,  M.-H.: Dimension formula for
graded Lie algebras and its applications. Trans. Amer. Math. Soc.
 {\bf 351},  4281-4336 (1999).
 
 \noindent 14.   Kang, S.-J.,  Kwon, J.-H.: Graded Lie superalgebras, supertrace formula and orbit Lie superalgebras. Proc. London Math. Soc.
 {\bf 81},  675-724 (2000).
 
\noindent 15. Landau, E.: Elementary number theory. AMS Chelsea Publishing (1966).

\noindent 16.  Lin,  M-S. M.: Applications of combinatorial analysis to the calculation of the partition function of the Ising model, Doctoral Dissertation, California Institute of Technology,  CA, USA, (2009).

\noindent 17.   Lis, M.: A short proof of the Kac-Ward formula. arXiv:1502.04322v1 [math.CO] (2015).

\noindent 18.  Loebl, M.: A discrete non-pfaffian approach to the Ising problem.
 DIMACS  {\bf 63}, Amer. Math. Soc., 
Providence, RI, 145-154 (2004). 

\noindent 19.  Markstr\"om, K.: The general graph homorphism polynomial:its relatioship with other graph polynomials and partition functions.  arXiv:1401.6335v1 [math.CO] (2014).

\noindent 20. Onsager, L.: Crystal Statistics. I. A two dimensional model with an order-disorder transition. Physical Review {\bf 65}, 117-149 (1944).

\noindent 21.  Serre, J. -P.: Lie algebras and Lie groups, Benjamin, New York, 1965.

\noindent 22.  Sherman, S.: Combinatorial aspects of the Ising model for
ferromagnetism.I. A conjecture of Feynman on paths and graphs. J. Math. Phys.
{\bf 1},  202-217  (1960).

\noindent 23.  Sherman,  S.: Combinatorial aspects of the Ising model for
ferromagnetism.II. An analogue to the Witt identity. Bull. Am. Math. Soc.
 {\bf 68}, 225-229 (1962).
 
\noindent 24.  Stark,  H. M.,  Terras, A. A.: Zeta Functions of finite graphs and coverings. Adv. Math. {\bf 121}, 124-165 (1996).

\end{document}